\documentclass[manuscript]{acmart}
\usepackage{graphicx}
\usepackage{multirow}
\usepackage{float}
\usepackage{threeparttable}
\usepackage{booktabs} 
\AtBeginDocument{%
  }

\setcopyright{acmlicensed}
\copyrightyear{2018}
\acmYear{2018}
\acmDOI{XXXXXXX.XXXXXXX}
\acmConference[Conference acronym 'XX]{Make sure to enter the correct
  conference title from your rights confirmation email}{June 03--05,
  2018}{Woodstock, NY}
\acmISBN{978-1-4503-XXXX-X/2018/06}

\author{Yibo Meng}
\affiliation{%
  \institution{Tsinghua University}
  \city{Beijing}
  \country{China}
}
\email{mengyb22@mails.tsinghua.edu.cn}

\author{Xiaolan Ding}
\affiliation{%
  \institution{North China University of Science and Technology, Health Science Center}
  \city{Tangshan}
  \country{China}
}

\author{Lyumanshan Ye}
\affiliation{%
  \institution{Shanghai Jiao Tong University}
  \city{Shanghai}
  \country{China}
}

\author{Zhiming Liu}
\affiliation{%
  \institution{University of Shanghai for Science and Technology}
  \city{Shanghai}
  \country{China}
}

\author{Yan Guan}
\affiliation{%
  \institution{Arts \& Design Academy, Tsinghua University}
  \city{Beijing}
  \country{China}
}
\email{guany@tsinghua.edu.cn}

\begin{document}

\title[A Longitudinal Study on the Attitudes of Gay Men in Beijing]{A Longitudinal Study on the Attitudes of Gay Men in Beijing Towards Gay Social Media Platforms: Lonely Souls in the Digital Concrete Jungle}

\renewcommand{\shortauthors}{Meng et al.}


\begin{abstract}
Over the past decade, specialized social networking applications have become a cornerstone of life for many gay men in \textcolor{black}{major metropolitan areas in China, such as Beijing}. This paper employs a longitudinal mixed-methods approach to investigate how \textcolor{black}{men in Beijing who have sex with men (MSM)} have shifted their attitudes toward these platforms between approximately \textcolor{black}{2014 and 2023}. Drawing on archival analysis of online discourses and \textcolor{black}{in-depth semi-structured interviews with 17 participants}, we trace the complex trajectory of this evolution. Our findings reveal a clear pattern: from the initial embrace of these applications as tools for community building and identity affirmation (2014--2017), to a period of growing ambivalence and critique centered on commercialization, ``hookup culture,'' and forms of discrimination (2017--2020), and finally to the present era (2020--2023), characterized by pragmatic, fragmented, yet simultaneously critical and reconstructive uses. Today, users strategically employ a repertoire of applications---including global platforms (e.g., Grindr and Tinder), domestic mainstream platforms (e.g., Blued), and niche alternatives (e.g., Aloha)---to fulfill differentiated needs. We develop a detailed temporal framework \textcolor{black}{theorizing from this urban context} to capture this attitudinal evolution and to discuss design implications for creating more supportive, secure, and community-oriented digital environments \textcolor{black}{for sexual minority users in China}. \textcolor{black}{While our findings are grounded in a Beijing-based longitudinal panel, Beijing represents an important urban environment where many digital practices among Chinese MSM have been pioneered and observed. Our theoretical claims should therefore be understood as emerging from this specific socio-urban context.}
\end{abstract}






\maketitle

\section{Introduction}

Digital dating applications such as Grindr\cite{spiel2019queer, taylor2017social, warner2018privacy, birnholtz2015let}, Tinder\cite{zytko2021computer, macleod2019construction}, and Blued\cite{boiano2008gender} have transformed intimate and community relationships among sexual minorities worldwide ~\cite{10.1145/3706598.3713813}. In China, since the mid-2010s, there has been a surge of social networking applications specifically targeting homosexual users. These platforms have provided gay men with unprecedented opportunities to express themselves, seek partners, and build communities in a society where the offline visibility of LGBTQ+ ~\cite{10.1145/3411763.3450403, 10.1145/3173574.3173881, 10.1145/3673229,10.1145/3613904.3642494, 10.1145/3274313} groups remains limited.

For marginalized communities\cite{hayes2020inclusive,davis2016digital,anuyah2023engaging,fox2016exploring,sannon2022privacy,rankin2020intersectionality,sin2021digital,erete2021can,olson2023along,devito2019social,schlesinger2017intersectional}, digital platforms are often regarded as \textit{``digital sanctuaries''} that offer connection\cite{nijs2020fostering}, recognition, and support. In the Chinese sociocultural context, where offline spaces are scarce, such platforms serve as vital social channels for sexual minorities (LGBTQ+)\cite{kannabiran2011hci, ahmed2018s, saha2019language, 10.1145/3555556, 10.1145/3170427.3170639, 10.1145/3025453.3025766, 10.1145/3334480.3381058}. While scholars generally acknowledge the empowering effects of these platforms, there is still a lack of in-depth longitudinal research on the long-term evolution and complexities of user experiences---particularly concerning the dynamic interactions between technology, users, and the broader social environment.

This study seeks to address that gap through a nine-year (2014--2023) mixed-method longitudinal investigation ~\cite{aalbers2022caught} that follows a cohort of Chinese gay men and analyzes the evolving relationships they maintain with specialized social applications~\cite{burke2010social,cheng2019understanding,chou2012they,cotter2024if}. The central argument of this research is that Chinese gay men’s relationship with same-sex social networking applications is not a linear trajectory of acceptance or rejection, but rather a dynamic process of evolution, which can be broadly divided into three stages:

\begin{itemize}
    \item \textbf{Early stage (2014--2017):} Characterized by high optimism and community building. Users generally regarded emerging applications as emancipatory tools, placing great hope in their capacity to break down information barriers, foster communities, and affirm identities.
    \item \textbf{Middle stage (2017--2020):} Marked by widespread disillusionment and critical reflection. As platforms became increasingly commercialized, community-level problems (e.g., discrimination, deception) grew more prominent, and state regulation tightened, users’ initial ideals were eroded, leading to prevailing attitudes of criticism and disappointment.
    \item \textbf{Recent stage (2020--2023):} Defined by pragmatic and strategic use. Users abandoned the expectation of finding a singular \textit{``digital home''} and instead adopted a practical, fragmented \textit{``multi-platform toolbox''} approach to meet diverse needs across different contexts.
\end{itemize}

\textcolor{black}{We adopt portfolio strategy as our primary explanatory lens: Chinese gay men continuously reconfigure combinations of apps in response to persistent sociocultural pressures, with platform affordances serving as contextual triggers rather than sole drivers of change~\cite{haimson2018social}.} This research examines how Chinese gay men’s attitudes toward such platforms have evolved over the past decade. Specifically, it addresses the following questions: 
\begin{enumerate}
    \item How have their perceptions of usability, privacy, community, and intimacy changed over time?
    \item What cultural, social, and technological factors have driven these changes?
    \item What insights can be drawn to inform the construction of more inclusive digital infrastructures?
\end{enumerate}

\textcolor{black}{By grounding a temporal account of queer digital adaptation in established theories of performative identity, privacy, and affordances, this study contributes a theoretically integrated understanding of how marginalized users negotiate sociotechnical constraints across time.} The study provides detailed empirical evidence for understanding how marginalized groups adapt to technology under conditions of high sociopolitical pressure. It contributes a unique longitudinal perspective to the field of Human--Computer Interaction (HCI) and offers valuable implications for designing more resilient and secure digital spaces.

\section{Related Work: \textcolor{black}{LGBTQ+ Identities, Sociotechnical Risks, and Platformized Sociality}}

The decade-long experience of Chinese gay men with social apps~\cite{kong2018gay, bardzell2011pleasure, ma2024evaluating, li2023we, xie2025futures}, as documented in this study, provides rich and unique empirical evidence for several core research areas in Human-Computer Interaction (HCI)~\cite{cui2022we, o2025finding, winchester2012use, cui2022so} and Computer-Supported Cooperative Work (CSCW)~\cite{bardzell2020join, wang2022gay}. The trajectory of user attitudes—from initial feelings of liberation to later critical reflection and current strategic use—not only corroborates existing theories but also offers new insights and challenges their applicability within marginalized communities and high-pressure political contexts. This section situates our findings within key theoretical frameworks in HCI to articulate the study's theoretical contributions to the field. \textcolor{black}{Early work on online identity highlights tensions between authenticity, privacy, and impression management in digital sociality~\cite{whitty2008revealing}. Building on this foundation, LGBTQ+ HCI scholarship demonstrates that platforms are central sites for identity work, disclosure, and visibility negotiation~\cite{blackwell2016lgbt, devito2018too, 10.1145/3274313, 10.1145/3432951, hardy2022lgbtq, 10.1145/3462204.3481786, 10.1145/3613904.3642482, 10.1145/3643834.3661564}. Social support emerges across diverse sociotechnical environments—from fandom communities~\cite{10.1145/3359256} to VR spaces~\cite{10.1145/3411763.3451673, 10.1145/3544548.3581530}, information activism in rural contexts~\cite{jonas2024better, 10.1145/3686991}, and identity-based governance in online communities~\cite{10.1145/3710900, 10.1145/3406522.3446040}. Yet LGBTQ+ participation online is shaped by moderation politics and algorithmic curation biases~\cite{10.1145/3479610, 10.1145/3613904.3641949, 10.1145/3613904.3642509, 10.1145/3706598.3713618}. These systems produce ``algorithmic exclusion''~\cite{10.1145/3432951}, particularly for trans and disabled creators~\cite{10.1145/3613904.3641949}, and reinforce homonormative aesthetics in curated feeds~\cite{10.1145/3706598.3713618}. Queer joy flourishes despite these harms~\cite{10.1145/3706598.3713592}, reflecting HCI calls to move beyond deficit framings~\cite{10.1145/3673229}. Research increasingly critiques commercial infrastructures that marginalize queer self-expression and desires for solidarity~\cite{10.1145/3706598.3713136}. Affordances of LGBTQ-specific and mainstream social platforms shape relational and dating practices~\cite{10.1145/3229434.3229460, yeo2016relationships, 10.1145/3613904.3642482}. Even non-dating platforms enable romantic seeking and community building~\cite{10.1145/3555189}. HCI scholarship conceptualizes these dynamics as cross-platform ``personal social media ecosystems''~\cite{10.1145/3274313}, where LGBTQ+ users tactically modulate visibility and boundary management across contexts~\cite{10.1145/3449162, blank2014new}. China-based research further highlights how queer users negotiate stigmatization and censorship in platformized ecosystems~\cite{10.1145/3491102.3517624, 10.1145/3555189, 10.1145/3706598.3714013}. Mainstream and LGBTQ-specific apps alike require strategic self-presentation to mitigate outing and harassment risks~\cite{10.1145/2930971.2930973}. These sociotechnical constraints shape both dating and identity development for Chinese queer communities. Taken together, this literature foregrounds a persistent tension: platforms can simultaneously afford connection, vulnerability, and sociotechnical adaptation. However, existing research is primarily cross-sectional and tends to focus on single platforms. Very little is known about how queer users’ attitudes and strategies evolve longitudinally under shifting sociocultural and platform conditions. We address this gap through a multi-wave study of Beijing gay men’s changing perceptions of gay social media platforms.}

\subsection{Managing Identity in Networked Publics: From Self-Presentation to Strategic Visibility}

Users' online identity construction is a central theme in HCI research. In this study, participants' journey from viewing social apps as a medium for "sincere self-expression" to feeling pressured by "the gaze and discipline of others" vividly illustrates the complexity of online identity management~\cite{audulv2023time, bayer2020social, braun2006using}.

Sociologist Erving Goffman's dramaturgical theory provides a classic framework for understanding online identity management\cite{Knoblauch2011}. He likens social interaction to a theatrical performance, where individuals manage others' impressions on the "front stage" through speech, appearance, and behavior, while being able to relax on the "backstage"\cite{bernstein2013quantifying, boyd2007social}. A social media profile is undoubtedly a carefully arranged "front stage," where users engage in continuous self-presentation by posting content, selecting avatars, and writing bios to shape an idealized self-image\cite{bowman2023using}. The "hope and liberation" experienced by participants in the early phase of this study (2014-2017) stemmed from these apps providing a new, seemingly safe "front stage" where they could perform a queer identity that was suppressed in their offline lives\cite{lieberman2020two}.

\textcolor{black}{Prior research on online dating has shown how dramaturgical self-presentation shapes romantic and sexual relationship development in networked environments~\cite{whitty2008revealing}, and how users often navigate a ``privacy paradox'' in which desires for connection coexist with concerns about surveillance and disclosure~\cite{blank2014new}. These insights help contextualize our participants' shifting negotiations between visibility, intimacy, and safety over time.}

However, the structural features of social media pose a significant challenge to traditional self-presentation. HCI researchers use the concept of "context collapse" to describe this dilemma: online platforms compress multiple, previously separate social circles (e.g., family~\cite{dworkin2018state}, friends~\cite{manago2015social}, colleagues~\cite{mccarthy2008context}, potential partners) into a single audience~\cite{bernstein2013quantifying}. This co-presence of multiple audiences makes it difficult for any single "performance" to satisfy everyone's expectations, leading to significant social risks and psychological pressure~\cite{wahlstrom2002influence, olson2003human}. For marginalized communities like LGBTQ+ individuals~\cite{taylor2024cruising}, the risk of context collapse is particularly severe, as an inappropriate self-disclosure could lead to being "outed" in family or professional settings, resulting in real-world discrimination and harm.

Therefore, for the participants in this study, online identity management~\cite{dimicco2007identity} is far more than simple impression management; it is a matter of survival through strategic visibility management. They must constantly weigh the benefits of visibility (e.g., finding community, gaining support, building intimate relationships~\cite{vetere2005mediating}) against the risks (e.g., harassment, discrimination, and political censorship). The longitudinal data from this study clearly show the dynamic nature of this trade-off: in the early stages, the benefits of visibility were widely favored, and users actively embraced the sense of liberation offered by the platforms. However, as negative experiences both on and off the platforms accumulated (such as the prevalence of discriminatory speech and the shutdown of Zank), the risks of visibility became increasingly prominent, prompting users to adopt more cautious and fragmented identity management strategies.

\subsection{The Shaping Role of Platform Affordances in Queer Sociality}

User social experiences are not created in a vacuum but are profoundly shaped by the characteristics of the technological tools they use. The theory of affordances in HCI~\cite{kaptelinin2012affordances} provides a precise analytical tool for explaining how platform design directly leads to the different social outcomes observed in this study. An affordance refers to the action possibilities~\cite{blin2016theory} that an object or environment offers to an actor~\cite{hartson2003cognitive}.

The different designs of the core apps in this study afforded distinct social possibilities:
\begin{itemize}
    \item \textcolor{black}{Myers~\citep{myers1994challenges}}'s grid interface afforded immediate encounters based on geographical proximity, greatly facilitating users seeking offline contact.
    \item Swipe-based matching mechanism~\cite{li2020swipe} afforded a connection model based on mutual selection, adding a layer of deliberation and consent to interactions.[7]
\item Disalvo's interest group feature afforded community\cite{disalvo2014making} building around shared hobbies rather than purely sexual attraction.
\end{itemize}
These different design choices provided users with a variety of ``digital homes"~\cite{desjardins2015investigating} in the early stages of the study, meeting their diverse needs from finding sexual partners to building friendships and communities. Similarly, platform design can also afford negative social interactions~\cite{lakey1994negative, rook1984negative}. The discrimination and harassment phenomena widely reported by users in the middle to late stages of this study (2017-2020) are also closely related to platform design choices. For example, many platforms set fixed role tags (e.g., ``1/0") in profiles, which, while convenient for quick filtering, also solidified and reinforced stereotypes and role discrimination within the community. Recommendation algorithms that overemphasize singular dimensions like appearance can afford body shaming and anxiety~\cite{schulenberg2023creepy}. And while platforms offer convenient anonymity and instant messaging, a lack of effective identity verification and community governance mechanisms can afford deceptive behavior and online harassment.

A core finding of this study—the continuous rise in user satisfaction with app usability (e.g., simple interface, reasonable functions) alongside a sharp decline in satisfaction with sociability outcomes (e.g., sense of belonging, long-term relationships)—can be profoundly explained by affordance theory. This phenomenon reveals a disconnect~\cite{nguyen2021managing} between usability affordances and sociability affordances in platform design. As platforms accelerated their commercialization, their design iterations increasingly focused on optimizing user retention and monetization, such as through algorithmic recommendations and live-streaming gifts. These improvements may have enhanced the operational fluency of certain interfaces (usability affordances) but did so at the expense of the community's healthy ecosystem, weakening the sociability affordances that promote deep connections and community trust. The results strongly demonstrate that for social platforms serving marginalized communities, technological ``progress" that contradicts the core social needs of the community will ultimately lead to user alienation and disillusionment. \textcolor{black}{Rather than fully determining user trajectories, these affordances act as situational triggers that interact with sociocultural pressures to shape evolving strategies of use.}

\subsection{Extending Social Transition Machinery to Longitudinal Identity Management}

The user behavior pattern observed in the later stage of this study (2020-2023)—strategically combining multiple domestic and international apps to meet different needs—is highly consistent with the theory of ``social transition machinery" proposed by HCI scholar Oliver Haimson~\cite{haimson2018social}. Furthermore, the longitudinal data from this study provide a unique perspective for extending this theory.

Haimson's framework suggests that individuals undergoing major life transitions (such as a queer gender transition) utilize multiple social media platforms with different functions and separate audiences, combining them into a coordinated ``machine" to manage the complex process of identity reconstruction. For example, one might use one platform for private identity~\cite{light2011hci} exploration and community support, while using another for public announcements to a broader social network.

The behavior of the participants in this study is a vivid illustration of this mechanism: they use Blued for location-based dating needs, Hinge or Tinder for deeper conversations or connecting with different social circles, and WeChat~\cite{chen2018understanding} as a channel for deepening trust and moving relationships offline. This fragmented usage strategy is a personalized digital ecosystem that users actively construct to navigate the limitations of a single mainstream platform and the significant risks of ``context collapse."

However, the decade-long tracking data of this study reveal a deeper implication than Haimson's framework~\cite{haimson2018social}. Haimson's theory primarily focuses on discrete life events with clear start and end points (e.g., a one-time ``coming out"). But the data from this study show that for marginalized groups living under continuous social pressure and uncertain regulatory environments, identity management is not a one-time ``transition" but a long-term, continuous state of adaptation and survival. The app portfolios they construct do not serve a temporary transitional period but have become a normalized ``continuous social adaptation machinery." Therefore, this study extends Haimson's theory from a tool for explaining discrete transitions to a framework for understanding the long-term resilience and digital survival practices of marginalized communities. Users' strategic behavior is not just an immediate reaction to platform limitations but a mature digital literacy evolved over a decade to cope with persistent social risks.

Finally, the findings of this study engage in a profound dialogue with the extensive literature in the HCI field on online support~\cite{liang2021embracing}, mental health~\cite{thieme2020machine}, and community building~\cite{dell2016ins} for marginalized communities. The results, from a longitudinal perspective, demonstrate the inherent "double-edged sword" nature of digital platforms for marginalized groups.

A large body of HCI research confirms that online platforms can provide crucial social support~\cite{esposito2015needs, wang2021cass, ta2020user}, information resources~\cite{wright2000analyzing}, and a sense of community for marginalized groups~\cite{liang2021embracing} isolated by geographical, social, or physical reasons. The early stage of this study (2014-2017) perfectly illustrates this, with participants hailing these apps as ``revolutionary tools" where they found ``a sense of belonging," received ``psychological support," and promoted ``self-acceptance". However, these supportive spaces can also become sources of immense psychological pressure. The mid-to-late stage data of this study (2017-2023) clearly document this shift: the former "safe havens" were gradually eroded by "deception," "discrimination," and "hookup culture," leading to a decline in some participants' self-worth, to the point where they began to doubt that "long-term same-sex relationships are inherently impossible".

The HCI field has consistently called for more longitudinal research~\cite{kjaerup2021longitudinal} to move beyond the limitations of cross-sectional studies and understand the dynamic, co-evolving relationship between technology and human society. This study directly answers that call. Through a decade-long tracking, the research not only demonstrates the coexistence of support and pressure but also reveals the dynamic process of their ebb and flow. 

In summary, this study provides the HCI and CSCW fields with a profound case study of the technology acceptance lifecycle for a marginalized community. It warns us that for designing for marginalized communities, merely providing the affordance of ``connection" is far from sufficient. If a design fails to actively and continuously commit to building an infrastructure of care—one that can resist commercial erosion, promote community health, and empower users to manage their own safety and well-being—then the initial promise of empowerment may eventually evolve into new forms of harm and exploitation. Digital platforms could then become venues that replicate and even amplify offline social injustices, rather than tools of liberation. This lays a solid theoretical and empirical foundation for the subsequent discussion of design implications with socio-cultural sensitivity.

\section{Method}

Over a period of nine years, we employed a mixed-methods approach, combining quantitative~\cite{adams2008qualititative} and \textcolor{black}{qualitative~\citep{mcdonald2019reliability, macleod2019construction}} research, to investigate the changing attitudes of 17 Chinese gay men towards gay social networking apps. The quantitative component utilized a usability scale to quantify participants' assessments of the basic viability of these apps.  The qualitative research consisted of semi-structured interviews covering: 1) basic user information (age, education, urban/rural residence, coming-out status); 2) types, frequency, and purposes of gay social media usage; 3) user experience with gay social apps, including functional design, interactive atmosphere, privacy and security, mental health, and social identity; and 4) open-ended questions about the limitations and potential development of existing apps, as well as expectations for digital technology, the future of the gay community, and their personal futures. 

\subsection{Participants}

Given the hidden nature of the target population (gay men who use same-sex social apps in China) and their strong overlap with digital media engagement, we recruited participants anonymously through widely used online platforms, including WeChat, Weibo, Baidu Tieba, as well as dedicated gay social networking apps such as \textit{Blued}, \textit{Fanka}, \textit{Aloha}, and \textit{Zank} (the latter of which has since been shut down by the Chinese government). We used a combination of random sampling and snowball sampling to maximize both representativeness and depth.

\begin{table}[htbp]
\centering
\caption{\textcolor{black}{Participant Information Across Four Waves of Interviews}}
\resizebox{\textwidth}{!}{
\begin{tabular}{cccccccccccccc}
\toprule
\multirow{2}{*}{ID} & \multirow{2}{*}{Age (2014)} & \multicolumn{3}{c}{2014} & \multicolumn{3}{c}{2017} & \multicolumn{3}{c}{2020} & \multicolumn{3}{c}{2023} \\
\cmidrule(lr){3-5} \cmidrule(lr){6-8} \cmidrule(lr){9-11} \cmidrule(lr){12-14}
 & & Education & Profession & U/R & Education & Profession & U/R & Education & Profession & U/R & Education & Profession & U/R \\
\midrule
1  & 19 & High school & Student & U & * & * & * & * & * & * & Bachelor & Official & * \\
2  & 19 & High school & Student & U & Bachelor & Secretary & U & Bachelor & Official & U & Bachelor & Official & U \\
3  & 22 & Bachelor & Student & U & * & * & * & Bachelor & Programmer & R & * & * & * \\
4  & 19 & High school & Student & U & Bachelor & Sex worker & U & Bachelor & Sex worker & U & Bachelor & Unemployed & U \\
5  & 24 & Bachelor & Chef & U & Bachelor & Chef & U & Bachelor & Chef & U & Bachelor & Chef & U \\
6  & 24 & Primary school & Chef & U & Primary school & Chef & U & Primary school & Chef & U & Primary school & Chef & U \\
7  & 27 & Bachelor & Labor & U & Bachelor & Labor & U & Bachelor & Labor & U & Bachelor & Labor & U \\
8  & 55 & Primary school & Farmer & R & Primary school & Farmer & R & Primary school & Farmer & R & Primary school & Farmer & R \\
9  & 33 & Bachelor & Sales & U & Bachelor & Sales & U & Bachelor & Deliveryman & U & Bachelor & Deliveryman & U \\
10 & 24 & Bachelor & Painter & U & Bachelor & Painter & U & Bachelor & Painter & U & Bachelor & Painter & R \\
11 & 44 & Bachelor & Labor & R & Bachelor & Labor & U & Bachelor & Labor & U & Bachelor & Unemployed & U \\
12 & 40 & Bachelor & Teacher & U & Bachelor & Teacher & U & Bachelor & Teacher & U & * & * & * \\
13 & 37 & Master & Student & U & PhD & Teacher & U & PhD & Unemployed & U & PhD & Unemployed & U \\
14 & 39 & PhD & Teacher & U & PhD & Teacher & U & PhD & Teacher & U & PhD & Unemployed & U \\
15 & 25 & High school & Deliveryman & U & High school & Deliveryman & U & High school & Deliveryman & U & High school & Labor & U \\
16 & 24 & Bachelor & Waiter & U & Bachelor & Waiter & U & Bachelor & Cleaning & U & * & * & * \\
17 & 24 & High school & Deliveryman & R & High school & -- & U & * & * & * & * & * & * \\
\bottomrule
\end{tabular}
}
\label{tab:participant_info}
\end{table}

\textbf{Inclusion criteria} were as follows: (1) aged 18 or older; (2) self-identifying as gay or questioning; (3) residing in China for at least five years, with familiarity with Chinese cultural norms and practices; (4) experience using gay social apps (e.g., \textit{Blued}, \textit{Aloha}, \textit{Fanka}, \textit{Zank}), or at minimum, an expressed interest in them; (5) basic language competence in Mandarin or a major Chinese dialect (e.g., Cantonese, Wu, Xiang, Gan, Jin); and (6) willingness to provide informed consent for long-term participation and follow-up.  

\textbf{Exclusion criteria} included: (1) under 18 years of age; (2) inability to communicate due to cognitive impairment or serious illness; (3) no prior experience with, or interest in, gay social apps; and (4) conflict of interest, meaning that researchers or their affiliates could not participate.  

\textbf{Ethics.} The study protocol was reviewed and approved by the Institutional Review Board (IRB) of an anonymized university. All participants were fully informed of the research objectives, procedures, and potential impacts. They were reminded of their rights to withdraw at any time without justification, to request deletion of their data, and to review any information collected about them. All data were anonymized, and pseudonymous participant IDs were assigned. Interviewers underwent extensive training in qualitative methods and LGBTQ+ cultural competence to ensure respectful engagement. Each participant received 30 RMB (approximately \$4.50 USD) for each completed interview session. \textcolor{black}{To better illustrate the longitudinal continuity and attrition in participants’ GSN app usage across the four data collection waves, Table \ref{tab:app_usage} summarizes which gay social applications each participant used in 2014, 2017, 2020, and 2023.}

\begin{table}[!h]
    \centering
    \small
    \caption{A summary of social applications \textcolor{black}{used by the same cohort of participants} used by participants across four years (2014, 2017, 2020, and 2023). The symbol `---' indicates that \textcolor{black}{missing data.}}
    \label{tab:app_usage}
    \begin{tabular}{cllll}
        \toprule
        \textbf{Participant} & \textbf{2014} & \textbf{2017} & \textbf{2020} & \textbf{2023} \\
        \midrule
        1 & Aloha, Blued, Finka & --- & --- & Ginder \\
        2 & Aloha, Blued, Finka & Aloha, Blued & Blued, Finka & Wechat, Tinder \\
        3 & Aloha, Blued, Finka & --- & Blued & --- \\
        4 & Aloha, Finka & Aloha, Finka & Blued, Aloha & Hinge, Blued \\
        5 & Aloha, Blued, Finka & Aloha, Blued & Blued & Hinge \\
        6 & Aloha, Blued, Finka & Aloha, Blued, Finak & Finka & QSmatch \\
        7 & Aloha, Blued, Finka & Aloha, Blued & Blued, Finka & QSmatch \\
        8 & Aloha, Finka & Aloha, Finka & Blued & Blued \\
        9 & Aloha, Blued, Finka & Blued & Aloha, Finka & Finka \\
        10 & Aloha, Finka & Aloha, Finka & Aloha, Finka & QSmatch \\
        11 & Aloha, Blued, Finka & Blued, Zank & Aloha, Blued & Blued \\
        12 & Aloha, Blued, Finka & Blued, Finka & Aloha, Finka & --- \\
        13 & Aloha, Blued & Aloha, Blued & Aloha, Finka & Blued \\
        14 & Aloha, Blued & Aloha, Blued & Aloha, Finka & Blued \\
        15 & Blued, Zank & Blued, Zank & Aloha, Finka, Tinder & Tinder, QSmatch \\
        16 & Blued & Blued, Finka & Blued, Aloha & --- \\
        17 & Aloha, Finka & Aloha, Finka & --- & --- \\
        \bottomrule
    \end{tabular}
\end{table}

\subsection{Procedure}

We conducted four rounds of data collection in April 2014, 2017, 2020, and 2023. Each round consisted of two components:  \textcolor{black}{\textbf{Step1:} Participants complete questionnaires to provide a baseline assessment of their perceived usability of gay male dating apps.} \textcolor{black}{\textbf{Step2:} Participants take part in a qualitative, semi-structured interview regarding their attitudes toward gay dating apps, lasting approximately 45–60 minutes.}

\subsection{Measures}

For the quantitative survey, participants rated 11 statements on a 10-point Likert scale. Items included, for example: ``I think the same-sex social app I use is simple and easy to use,” ``Same-sex social apps help me better accept my sexual orientation,” and ``Overall, I feel that same-sex social apps have a positive impact on me.” These items were designed to capture both functional evaluations and broader psychosocial perceptions.

The qualitative interview protocol consisted of structured domains, including: (1) demographic information (age, education, urban/rural background, coming-out status); (2) app usage patterns (platforms used, frequency, purposes); (3) user experiences (usability, design, privacy, discrimination, mental health, social identity); and (4) open-ended reflections on limitations of current apps, future expectations for technology, and broader life aspirations.  

\begin{table}[htbp]
\centering
\small
\caption{Attitude Items for Gay Social Networking Apps (GSN apps)}
\begin{tabular}{cl}
\toprule
ID & Item \\
\midrule
1 & I believe the GSN apps I use have a clean, usable design. \\
2 & Overall, GSN apps safeguard my privacy. \\
3 & The GSN apps I use offer a coherent feature set. \\
4 & I use GSN apps frequently. \\
5 & I like GSN apps (e.g., \textit{Zank}, \textit{Blued}, \textit{Aloha}). \\
6 & I believe GSN apps support my mental health. \\
7 & GSN apps help me accept my sexual orientation. \\
8 & GSN apps help me feel a sense of belonging. \\
9 & I find GSN apps more effective than offline socializing. \\
10 & I believe GSN apps facilitate forming long-term relationships. \\
11 & Overall, GSN apps have a positive impact on me. \\
\bottomrule
\end{tabular}
\label{tab:gsn_attitudes}
\end{table}

\noindent \textbf{Qualitative component.} Second, we drew on participants’ quantitative questionnaire results from this study to inform the interviews. We conducted semi-structured interviews. We further examined Chinese gay men’s specific attitudes toward gay social networking apps (GSN apps), the underlying reasons, and their recommendations. The detailed interview structure is as follows:
\begin{enumerate}
  \item demographics (age, education, urban--rural background, outness);
  \item types, frequency, and purposes of gay social media use;
  \item experiences with GSN apps, including feature design, interaction climate, privacy and safety, mental health, and social identity;
  \item development-oriented questions, including limitations of current GSN apps, directions for development, expectations for digital technology and for the future of gay men, and expectations for themselves.
\end{enumerate}

\begin{table}[htbp]
\centering
\caption{Interview Guide for GSN App Users}
\small
\begin{tabular}{p{3cm} p{11cm}}
\toprule
Category & Questions \\
\midrule
\textbf{Demographics} & 
1. What is your age? \newline
2. Do you currently live in an urban or rural area? \newline
3. What is your highest level of education? \newline
4. What is your coming-out status? \newline
5. What is your occupation? \\
\midrule
\textbf{GSN App Use} &
6. Which gay social networking (GSN) apps do you currently use or have used in the past? \newline
7. Have you ever used multiple GSN apps at the same time? \newline
8. How frequently do you use such apps? \newline
9. What are your main purposes for using these apps? \\
\midrule
\textbf{GSN App Experience} &
10. Do you think the interface design of the GSN apps you use is aesthetic and easy to navigate? \newline
11. What features do these GSN apps provide, and do you consider them well designed? \newline
12. Have the GSN apps you use fulfilled your intended purposes? How was your experience? \newline
13. Do you believe the social atmosphere on these apps is welcoming and users are friendly? \newline
14. Do you think you can meet genuine people on these apps? \newline
15. Have you had any negative experiences on GSN apps? \newline
16. Do you believe users on GSN apps place excessive emphasis on appearance, such as body shape or looks? \newline
17. Have you experienced body shaming on GSN apps? \newline
18. Have you encountered hookup culture on these apps, and did it make you uncomfortable? \newline
19. Have you witnessed discriminatory comments based on race, body type, or age? \newline
20. Have you experienced any other forms of discrimination on these apps? \newline
21. While using GSN apps, are you concerned about the leakage of location or other private information? \newline
22. Are you satisfied with the anonymity and privacy protection measures of these apps? \newline
23. Are you concerned about fraud, blackmail, or financial traps when using GSN apps? \newline
24. Have you encountered fake photos or false identity information on GSN apps? \newline
25. Do you believe using GSN apps helps you better accept your sexual orientation? \newline
26. Do you believe GSN apps help reduce anxiety or feelings of loneliness? \newline
27. Do you sometimes doubt yourself when you do not receive replies or matches? \newline
28. Are GSN apps an important channel for you to access LGBTQ+ related information? \newline
29. Do you think GSN apps provide personal support for you? \newline
30. Overall, how do you perceive GSN apps? \\
\midrule
\textbf{Open Questions} &
31. In your view, what are the greatest strengths of current mainstream GSN apps, and what weaknesses most need improvement? \newline
32. What would your ideal GSN app look like? What features or qualities should it include? \newline
33. Do you have any other thoughts, feelings, or suggestions you would like to share with us? \\
\bottomrule
\end{tabular}
\label{tab:gsn_interview}
\end{table}

\noindent \textbf{Interview procedure.}  
All interviewers received professional training in interviewing skills and basic knowledge of gay culture before the study. This ensured full respect for participants and allowed the collection of rich and in-depth information. The 2014 and 2017 interviews were conducted via online audio calls, while the 2020 and 2023 interviews were conducted via Tencent Meeting. All four rounds of interviews were audio-recorded. Participants were advised to use a voice changer to protect their vocal identity. Within 24 hours after each interview, the recordings were transcribed into text. Each transcript was checked and revised by at least two interviewers.

\subsection{Data Analysis}

Quantitative data were analyzed using descriptive statistics and exploratory factor analysis (EFA) in SPSS to identify attitude clusters. \textcolor{black}{Survey results across the four time points were compared to examine longitudinal shifts.} Qualitative data were transcribed and analyzed using thematic analysis, \textcolor{black}{following Braun and Clarke’s six-phase framework}. Coding followed an iterative process, \textcolor{black}{our qualitative analysis was guided by the contextual integrity framework~\cite{grimm2010social, nissenbaum2004privacy, nissenbaum2009privacy}}. According
to the framework, starting with open coding and then grouping by themes. \textcolor{black}{Two trained researchers independently conducted line-by-line coding in NVivo 12 using a hybrid deductive–inductive strategy, where initial codes were informed by the research aims and latent codes emerged from the data. Inter-rater reliability was assessed on 20\% of transcripts using Cohen’s Kappa ($\kappa = 0.86$), indicating strong agreement, and discrepancies were resolved through consensus discussions. Data saturation was monitored in each wave and typically reached after 10--12 interviews.} \textcolor{black}{To track thematic evolution diachronically, we calculated code frequencies for each theme across 2014, 2017, 2020, and 2023. Pearson’s Chi-square tests ($\chi^2$) were used to determine the significance of thematic shifts, supporting the triangulation of qualitative results with the quantitative findings.}

% \subsection{Data Analysis}  
% Quantitative data were analyzed using descriptive statistics and exploratory factor analysis (EFA) in SPSS to identify attitude clusters. Qualitative data were transcribed and analyzed using thematic analysis. Coding followed an iterative process, starting with open coding and then grouping by themes.  

\section{Results}  
In 2014, we recruited 17 participants, aged 19–55 ($M=29.35$, $SD=10.29$). Fourteen participants came from urban areas, and three from rural areas. Most participants had relatively high levels of education: ten held a bachelor’s degree or above, five held a bachelor’s degree (three of whom later confirmed obtaining this degree in follow-up interviews), and only two participants had completed only primary school. None were illiterate. In 2017 and 2020, two participants were absent in each round. In 2023, four participants were absent, leaving 13 participants. Their occupations were diverse, including sex workers, students, teachers, officials, laborers, couriers, farmers, and unemployed individuals. Detailed demographic information is shown in the table below.

\subsection{Quantitative Results}

\begin{figure}[!h]
    \centering
    \includegraphics[width=0.5\linewidth]{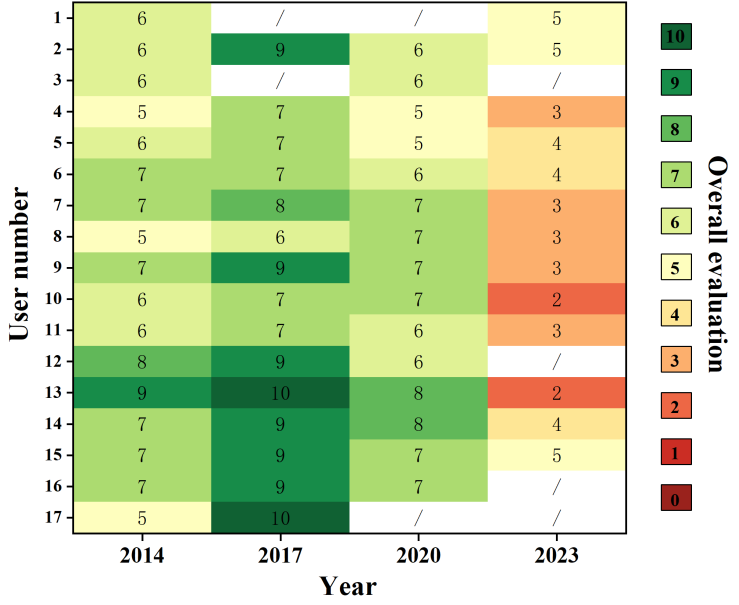}
    \caption{Overall evaluation of peer social apps from 2014 to 2023. The heatmap (top) shows individual user ratings across four time points, while the bar chart (bottom) presents mean overall evaluation with standard deviations. Results indicate fluctuations in user perceptions over the nine-year period.}
    \label{fig:placeholder}
\end{figure}

\begin{figure}[htbp]
\centering
\begin{minipage}{0.48\linewidth}
    \centering
    \includegraphics[width=\linewidth]{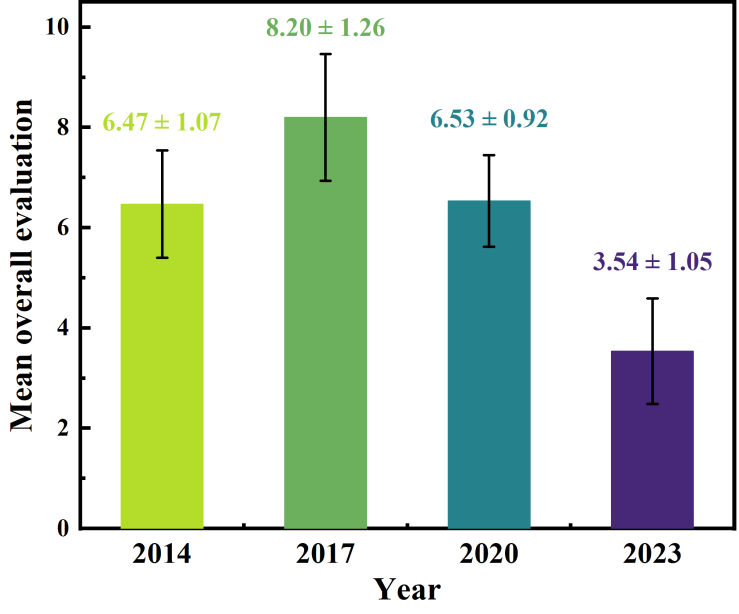}
    \caption{Mean overall evaluation of peer social apps from 2014 to 2023, with error bars indicating standard deviations.}
    \label{fig:figure2}
\end{minipage}\hfill
\begin{minipage}{0.48\linewidth}
    \centering
    \includegraphics[width=\linewidth]{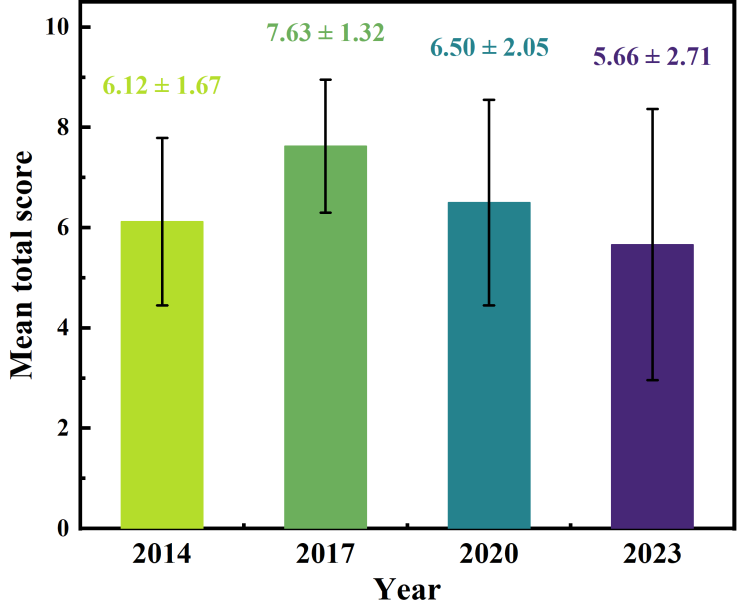}
    \caption{Mean total score of peer social apps across the same period, with error bars indicating standard deviations.}
    \label{fig:figure3}
\end{minipage}
\end{figure}

The data indicate two phases in overall attitudes toward gay social networking apps (GSN apps) from 2014 to 2023. From 2014 to 2017, the mean overall evaluation rose markedly. It increased from 6.47 ($SD=1.07$) in 2014 to 8.20 ($SD=1.26$) in 2017. The sizable gain and the small SDs suggest broadly positive evaluations among the 17 participants during this period. From 2017 to 2023, the mean declined steadily. In 2020, it was 6.53 ($SD=0.92$), only slightly above the 2014 baseline of 6.47. By 2023, it fell to 3.54 ($SD=1.05$), well below the 2014 level. The small SD again indicates a widely shared disappointment with GSN apps among participants.

Given potential bias in users’ subjective perceptions, we averaged all item scores for each of the four waves. The results closely matched participants’ subjective overall ratings. The mean across all 11 scale items rose from 6.12 ($SD=1.67$) in 2014 to 7.57 ($SD=1.33$) in 2017, then declined to 6.47 ($SD=2.05$) in 2020 and to 5.65 ($SD=2.70$) in 2023. The 2023 mean was clearly below the 2014 baseline. The data reveal a three-phase trajectory in attitudes toward gay social networking apps (GSN apps): an initial rapid increase (2014–2017), a peak, and a sustained decline (2017–2023).

\begin{figure}
    \centering
    \includegraphics[width=1\linewidth]{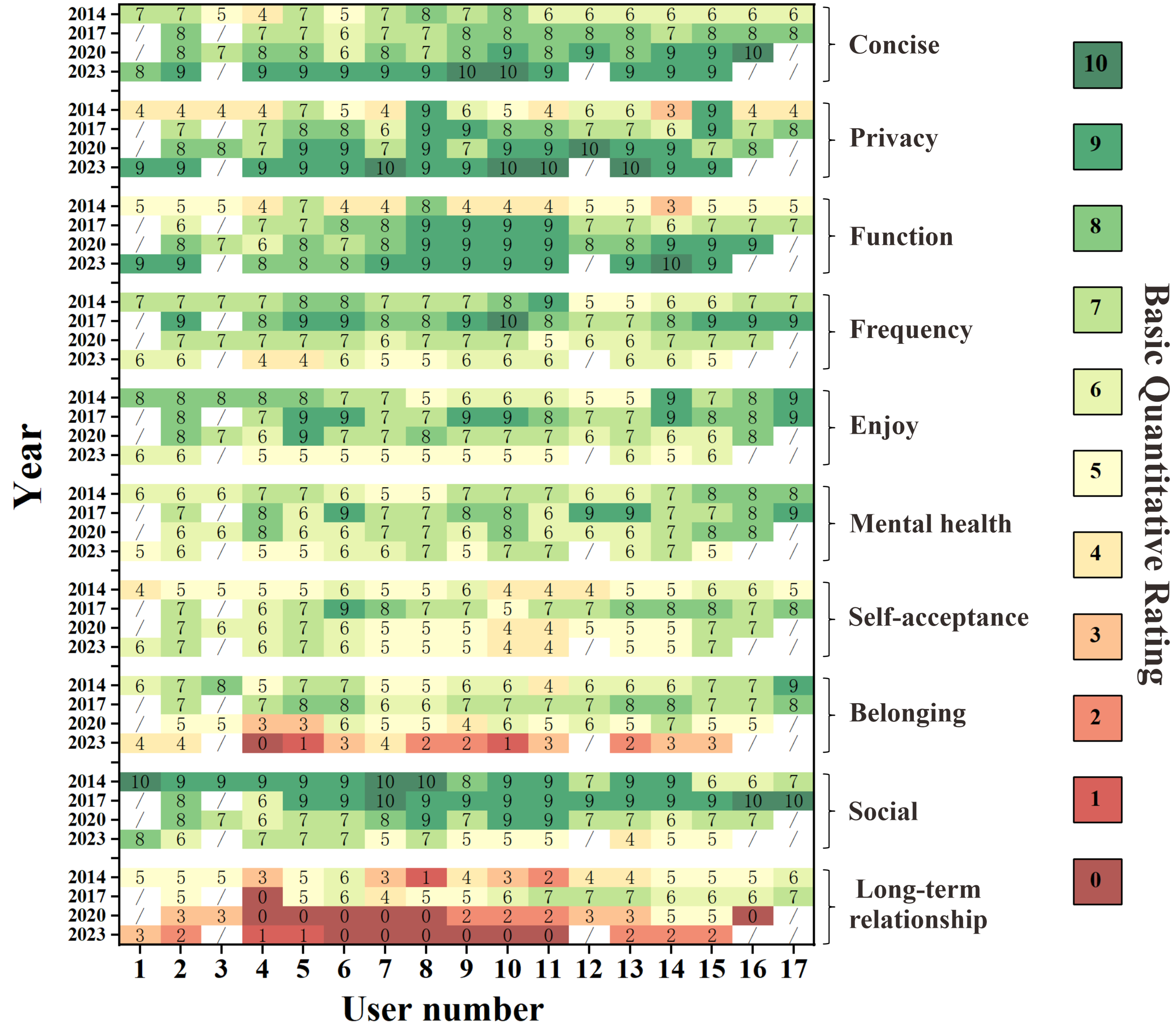}
    \caption{User evaluations of peer social apps from 2014 to 2023 across ten key dimensions: Concise, Privacy, Function, Frequency, Enjoy, Mental health, Self-acceptance, Belonging, Social, and Long-term relationship. Each row represents one dimension, while each column corresponds to a user. Ratings are color-coded from 0 (lowest) to 10 (highest).}
    \label{fig:placeholder}
\end{figure}

\begin{figure}
    \centering
    \includegraphics[width=1\linewidth]{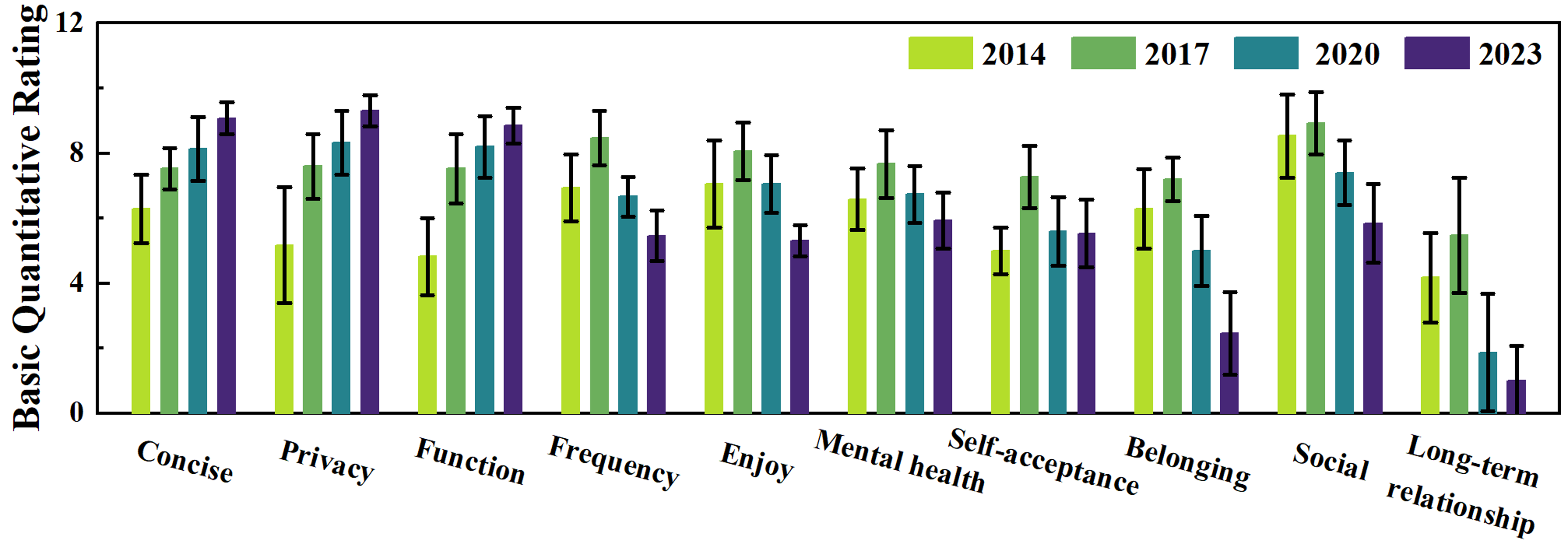}
    \caption{Mean user ratings of peer social apps across ten evaluation dimensions (Concise, Privacy, Function, Frequency, Enjoy, Mental health, Self-acceptance, Belonging, Social, and Long-term relationship) from 2014 to 2023, with error bars indicating standard deviations.}
    \label{fig:placeholder}
\end{figure}

Additionally, item-level trends show steady gains in basic performance ratings for gay social networking apps (GSN apps) from 2014 to 2023. The increases concern three dimensions: functionality appropriateness and completeness, clean and easy-to-use interface design, and protection of privacy. For example, the mean rating for ``clean interface” rose from 6.29 ($SD=1.05$) to 7.53 ($SD=0.64$), 8.13 ($SD=0.99$), and 9.08 ($SD=0.49$) across the four waves. Ratings for ``privacy” and ``functionality” followed the same upward pattern and increased throughout 2014–2023. These trends indicate ongoing improvements in basic capabilities: interactions became simpler; feature design was perceived as increasingly coherent; privacy was seen as better safeguarded. 

This trajectory diverged from the overall evaluation. After 2017, overall attitudes declined, yet recognition of basic functionality, interface design, and privacy protection continued to rise. The divergence suggests two points. First, from 2014 to 2023, app technology and core interface design improved and were acknowledged by users. Second, new user needs emerged over time that GSN apps did not address. In short, technical progress did not fully translate into user approval.

In contrast, the dimensions of ``frequency,” ``liking,” ``self-acceptance,” ``mental health,” ``belonging,” ``social interaction,” ``long-term relationship,” and the ``overall evaluation” followed a different trajectory. All increased markedly between 2014 and 2017, peaked in 2017, and then declined steadily through 2023 without exception. This pattern highlights the gap between users’ subjective experiences and the continuous technical improvement of GSN apps.  

The ``long-term relationship” item is most illustrative. Its mean rose from 4.18 ($SD=1.38$) in 2014 to a peak of 5.47 ($SD=1.77$) in 2017, but then declined sharply to 1.87 ($SD=1.81$) in 2020 and to 1.00 ($SD=1.08$) in 2023. It consistently had the lowest mean score among all items. Similarly, the ``belonging” item peaked at 7.20 ($SD=0.68$) in 2017 and then dropped rapidly, reaching 2.46 ($SD=1.27$) in 2023.  

The ``social interaction” item showed consistently higher scores. Except in 2023, when its mean fell below that of ``mental health,” it remained among the highest-rated items. This indicates that the basic social function of GSN apps was relatively well recognized by users, in contrast to higher-level functions such as ``long-term relationship,” ``self-acceptance,” and ``belonging.”

\subsection{Qualitative Results}

The qualitative findings largely corroborate the quantitative results. Initially, participants viewed gay social networking apps (GSN apps) as revolutionary tools for community building and identity affirmation. As a result, the apps were warmly received (2014–2017). From 2017 to 2020, technical architectures continued to improve. Basic usability also improved. Yet participants increasingly expressed disapproval on multiple fronts: commercialization; ``fast-food” interactions; hookup culture; lookism; role-based discrimination; and pervasive deceptive practices. Reliance on GSN apps began to decline during this period. In the 2020–2023 interviews, disappointment became salient. Participants displayed clearer segmentation and a pragmatic stance. Many adopted international GSN apps (e.g., \textit{Grindr}, \textit{Tinder}). Use of domestic mainstream apps (e.g., \textit{Blued}, \textit{Aloha}) decreased. \textcolor{black}{To ensure conceptual continuity across all four waves, we maintained a stable core codebook grounded in the research questions and prior literature. Two researchers re-coded a shared subset of earlier transcripts at each new wave to recalibrate interpretations of key concepts and prevent code drift. Any emergent refinements were applied retroactively to previous waves, ensuring that observed changes were due to genuine shifts in user experience rather than evolving coding criteria. Accordingly, $\chi^2$ statistics are used purely as descriptive indicators of theme prevalence under a consistent conceptual definition.}

\subsubsection{Hope and Liberation Afforded by Digital Sociality}

Overall, interviews in 2014 and 2017 revealed highly positive attitudes toward gay social networking apps (GSN apps). Favorability increased from 2014 to 2017. The interviews suggest four drivers of endorsement: self-expression, social connection, identity affirmation, and hope with development.

First, since 2014, the proliferation of GSN apps provided Chinese gay men with a medium for sincere self-expression, without facing suspicion or humiliation. Thirteen participants in 2014 and fifteen in 2017 mentioned this view or expressed similar ideas, for example:

\begin{quote}\itshape
On \textit{Blued}, I can say I like big pecs (a masculine trait). I do not have to pretend to like women’s bodies. In my high school dorm, some questioned my sexual orientation. When my roommates told explicit jokes about women, they would shame me.
\par\normalfont\hfill (Participant 1, 2014)
\end{quote}

\begin{quote}\itshape
I am the father of three children. They are all adults. In everyday life, I must perform as a good husband, a father, and a good son who has fulfilled patrilineal duty in Chinese culture. It exhausts me. Without \textit{Finka}, I would have no one to tell these things to, and I could not say them at all.
\par\normalfont\hfill (Participant 8, 2014)
\end{quote}

\begin{quote}\itshape
Society always dictates what a man should be. On \textit{Aloha}, it does not. I can say what I want\ldots
\par\normalfont\hfill (Participant 7, 2017; excerpt contains explicit male sexual content; portion omitted)
\end{quote}

\begin{quote}\itshape
This is thrilling. I open the app and see hundreds of men like me. I am no longer alone.
\par\normalfont\hfill (Participant 17, 2017)
\end{quote}

The interview cases above suggest the following. In the formative stage of Chinese gay social networking apps (GSN apps) from 2014--2017, these apps offered a platform for expressing sexual orientation and preferences. The platform operated outside the conventional cultural order. Potential issues also emerged. Sexually explicit content could appear on these platforms. Some participants reported engaging in extramarital behavior within heterosexual marriages. Nevertheless, participants commonly stated that they had found a window for free self-expression without fear of censure.

Second, social connection. GSN apps gathered gay men from different corners of China who were ``hidden'' among those perceived as ``normal'' in the Chinese cultural context. The platforms enabled the search for peers, friends, sexual partners, and romantic partners. All participants in 2014 and 2017 mentioned this aspect. For example:

\begin{quote}
I do have many friends. Yet I could never tell them that last year I was preparing to come out to my family. Our relationships are good. But they often insult gay people as ``disgusting” in front of me. Coming out brought immense pressure. My family did not support me and has almost cut off my financial support. Before I encountered Blued, I had nowhere to voice these grievances. Speaking up rashly would likely invite another round of ridicule. (Participant 16, 2014)
\end{quote}

\begin{quote}
We often talk for hours about the type of men we like, whether our families are supportive, and our future plans. We meet for coffee, simply to make friends. (Participant 4, 2014)
\end{quote}

\begin{quote}
I found my sexual partner through Aloha. We get along very well. You cannot infer someone’s sexual orientation from the venue (here referring to other social platforms) or from face-to-face interactions offline. (Participant 8, 2017)
\end{quote}

\begin{quote}
I met my boyfriend on Zank. We later registered our marriage in Canada. It all feels like a dream, doesn’t it? (Participant 15, 2017)
\end{quote}

The foregoing accounts indicate that gay social networking apps (GSN apps) reinforced multiple layers of social ties: friendship, romantic relationships, and sexual contacts. Similar descriptions recurred with high frequency in interviews conducted in 2014 and 2017. In 2014 alone, across transcripts from only 17 participants, the word ``friend” appeared 322 times; ``lover,” ``partner,” or ``boyfriend” appeared 341 times; and ``hook-up” or ``sexual partner” appeared 67 times. There is little doubt that GSN apps expanded the social networks of gay men in China.

During the same period, GSN apps also strengthened participants’ self-identity. This aligns with earlier observations on fluent self-expression and the construction of rich social ties. The effect manifested as affirmation of sexual orientation, of sexual behaviors, and even of self-worth. For example:

\begin{quote}
Everyone is different. Even for someone like me (a sexual minority), our existence is equally legitimate. As the lyric goes, ``I am what I am, a firework of a different color.” This is the key insight I gained from gay social networking apps. I no longer feel the need to prove anything to others. There is nothing more to explain. (Participant 3, 2014)
\end{quote}

\begin{quote}
No one in my class paints male nudes (a common oil-painting subject) better than I do. I posted these works on Aloha. Many users liked them. I am proud. I now update my painting feed nearly every few days. (Participant 10, 2014)
\end{quote}

\begin{quote}
Do you not think so? This is not boasting. Look at the posts on Finka. We gay men are highly accomplished. The share with higher education and higher income among us far exceeds that among so-called ``normal people.” (Participant 10, 2017)
\end{quote}

Finally, we observe growth-oriented aspirations. Gay social networking apps (GSN apps) provided some idealistic participants with a platform for collaboration and collective advancement. The platform supported their pursuit of, and efforts toward, an envisioned better world. For example:

\begin{quote}
Equality across sexual orientations will come, sooner or later, so long as we remain united and brave. (Participant 4, 2017)
\end{quote}

\begin{quote}
I founded a gay community at my university. We recruit new members via Blued, Aloha, and Finka. We protest the current ``tyranny” together. (Participant 2, 2017)
\end{quote}

This phenomenon was more salient among younger participants. University students with higher education articulated clear ideals, including how they believed the world should be and how it should regard the gay community. Gay social networking apps (GSN apps) facilitated their connection, gathering, communication, and even expansion.  

For these reasons, participants in the 2017 interviews expressed the strongest recognition of GSN apps. When asked about their overall views or feelings, almost no negative expressions appeared. Only Participants 4, 5, and 7 raised concerns about the difficulty of finding long-term relationships. This aligns with quantitative findings, which showed that participants’ overall evaluation of GSN apps peaked in 2017.

\subsubsection{Advances in technology, and the rise of tensions and critique}

When discussing core attributes—app functionality, interface design, and privacy/security—participants’ satisfaction continued to increase. In the 2020 interviews, almost no one raised concerns about privacy breaches. By contrast, the issue appeared 11 times in 2014 and 7 times in 2017. For example:

\begin{quote}
What if people in my village find out that I am gay? I am afraid of matching with acquaintances nearby. (Participant 8, 2014)
\end{quote}

\begin{quote}
What if others save my photos and misuse them? Although I am an early adopter, I still worry. Even if a leak happened, it might not be a big problem—who would target me? But it would force me to come out. (Participant 1, 2014)
\end{quote}

These concerns did not appear in the 2020 interviews. Participants noted that most GSN apps—such as Blued, Aloha, and Finka—had implemented anonymity, location obfuscation, profile and message controls, and disappearing photos (auto-deleted after viewing). These features enabled confident use.

However, in contrast to technical advances, from 2017 to 2020 participants’ discourse about GSN apps became increasingly negative, consistent with the quantitative findings. The main factors included:
\begin{itemize}
\item pervasive deception;
\item widespread discrimination (e.g., looks, education, age, and sexual-role discrimination);
\item the prevalence of fast-paced dating and hook-up culture;
\item excessive commercialization.
\end{itemize}

First, deception and falsehoods were common on GSN apps. Participants reported diverse forms of lying, for example:

\begin{quote}
Blued allows anonymous personal information and does not verify users’ actual age. You can write whatever you want. I once chatted for nearly half a year with someone who claimed to be 28. He asked me out for dinner, and I… (explicit content omitted). But I swear, he must have been at least 50. He really… (offensive language omitted). I was furious! (Participant 2, 2020)
\end{quote}

\begin{quote}
I was involved with an official for about two months. We chatted continuously on Aloha. He told me he was looking for a partner, and I said I was too. Later, we met and had sex. After that, he never contacted me again. I later heard from his friends that he was chatting with at least seven or eight other ``boyfriends” on Aloha at the same time. I have to say, Aloha is really ``convenient” (a sarcastic remark on his use of Aloha to date multiple people simultaneously). (Participant 9, 2020)
\end{quote}

\begin{quote}
On Finka he told me he was a ``1” (a term in Chinese gay contexts referring to the insertive role during sex). But when we met, he said he was only a ``0” (referring to the receptive role). You know, I was speechless. Couldn’t he clarify this in advance? Why did he have to lie? (Participant 14, 2020)
\end{quote}

\begin{quote}
People say anything online. Just listen and let it pass. I even came across one of my lecturers claiming that he graduated from Tsinghua University (a prestigious university in China). In fact, he did not know I was his student. I knew he actually graduated from a second-tier university. (Participant 15, 2020)
\end{quote}

\begin{quote}
He said on Blued that he had never had sex. Personally, I do not think sex itself is wrong. But later, I found out that his… you know. I definitely did not believe he had never had sex. (The participant withheld details but at least conveyed his distrust of his partner’s statement.) (Participant 16, 2020)
\end{quote}

Participants reported that gay social networking apps (GSN apps) reproduced core social-media affordances: a sense of unreality, anonymity, and low barriers to participation. Deception on these platforms spanned most domains of life, including, but not limited to, age, height, sexual history, sexual role, relationship intentions, assets, authority or status, and occupation. Although individual accounts may be limited by perspective, distrust toward information exchanged on GSN apps was evident.

Second, discrimination was likewise commonplace. For example:
\begin{quote}
To be honest, I do not find gay social networking apps particularly useful. At times, I even hate them, though I still use them. I am unattractive—or at least many people think so. I am not afraid you will laugh. Each time I start a chat, after a line or two the other person says, ``let me see you.” I send my photo. The conversation then ends. (Participant 7, 2020)
\end{quote}

\begin{quote}
You can see I have put on some weight. I remember the first time I posted a photo on Aloha. There were at least twenty comments calling me a ``pig.” (Participant 11, 2020)
\end{quote}

\begin{quote}
On those apps, anything can become a basis for discrimination. You even see routine discrimination based on the appearance of one’s genitals. These things are hard to say aloud. Sometimes, when I use those apps, I feel ashamed of being gay. (Participant 2, 2020)
\end{quote}

Participants 7, 11, and 2 raised a representative concern. Among the 15 participants interviewed in 2020, nine reported experiencing appearance-based discrimination on GSN apps, or at least feeling psychological pressure. Among the remaining six, four stated that such discrimination was reasonable, or acknowledged having discriminated against others in similar ways.

Role-based discrimination was likewise commonplace on GSN apps. For example:

\begin{quote}
Let me tell you something interesting. For a while I was curious about the other role. On Aloha I changed my role from ``0” (receptive) to ``1” (insertive). My followers jumped by more than 300 in a month. Over 120 people messaged me. Later I was no longer curious and switched back to ``0.” About 50 messages insulted me. (Participant 4, 2020)
\end{quote}

\begin{quote}
When browsing GSN apps, I sense a valorization of the insertive role in our culture. Even as a gay man, other gay men ``expect” me to be sufficiently ``manly” (a stereotyped form of masculinity). (Participant 13, 2020)
\end{quote}

In addition, age-based and education-based discrimination was common. Eleven participants reported experiences of education shaming (based on academic credentials). Five participants reported age shaming. For example:

\begin{quote}
There are all kinds of people on Blued. I am a sex worker. Once, after finishing work, a middle-aged client began lecturing me about not studying hard at my young age. I was speechless. Why should it matter to him if I left school? I completed compulsory education (nine years of state-mandated schooling in China, up to the end of junior high). Whether I attend university or not is my decision. (Participant 4, 2020)
\end{quote}

\begin{quote}
If you log on to Finka, or the now-banned Zank, you will see countless users writing ``Well-educated” in their profiles. There is little more to say. People usually shout loudest about what they actually lack. (Here the participant used a Chinese colloquialism to satirize the lack of genuine educational quality among those who look down on less-educated groups.) (Participant 2, 2020)
\end{quote}

\begin{quote}
I am not someone who likes to lie. I listed my real age on all my app profiles. I thought this was a responsible act toward others. But I did receive many offensive remarks. Once I posted a photo of myself working out. The first comment said: ``You look strong, but unfortunately at 45 your sexual performance must be failing.” (The participant’s account included sexual language, paraphrased here.) I felt deeply offended. (Participant 14, 2020)
\end{quote}

As gay social networking apps (GSN apps) evolved, their user base expanded, and users became more familiar with them. In interviews, the frequency of the above behaviors rose markedly. Reports of education shaming and age shaming became more salient. Participants noted that social apps do more than express users’ own agency. The gaze and disciplining of others are also mediated through these apps and intrude pervasively on users’ lives.

In addition, the prevalence of fast-paced dating and hook-up culture was a major reason for declining interest in GSN apps. In the digital era, interpersonal relations are continually deconstructed. Ties grow fragile. Relationships tend to be short-term and high-frequency. GSN apps amplify this trend, for example:

\begin{quote}
On Aloha there are no real relationships, only sexual ones. People chat almost exclusively for sex. I feel as if I am not a person, but a ``cup” (a sex toy, here expressing the participant’s frustration at being treated merely as a sexual object). (Participant 6, 2020)
\end{quote}

\begin{quote}
You have never been in our circle, have you? In this circle on gay social apps, it is ``a battlefield of hookups” (a metaphor for chaotic sexual relations). Anything can happen. Do not even mention love; sometimes there is no morality at all. I once dated a ``0” (receptive role). I treated him as a partner, and we chatted for a very long time. During one of our encounters, his wife called. Only then did I learn he had long been married, and his wife was nine months pregnant and in the delivery room. You may be surprised, but do not be. There are countless stories here you would not expect. (Participant 14, 2020)
\end{quote}

\begin{quote}
In these apps, if a relationship lasts one month it is considered romantic, two months counts as ``old couple,” and if it survives three months without breaking up, it is seen as ``a gift from heaven” (a sarcastic remark on the brevity of gay relationships). (Participant 7, 2020)
\end{quote}

The above participants expressed aversion to short, fast-paced ``hook-up culture.” This aligns with quantitative findings: after 2017, GSN apps increasingly failed to provide ``long-term relationships,” ``sociality,” or ``belonging.” The pattern stands in sharp contrast to the frequently cited themes in 2014 and 2017 of ``building a future” and ``an equal world.” It signals the collapse of an earlier ``online utopia” envisioned by some participants.

Finally, participants noted an atmosphere of excessive commercialization. This issue was less prominent; three participants mentioned it. Manifestations included paywalls for app functions and an overabundance of advertising.

\subsubsection{Pragmatism and Development}
At this stage, participants’ overall evaluation of GSN apps reached its lowest point. However, compared with 2020—when critiques were largely negative—more pragmatic and constructive expressions emerged. At the same time, clear differentiation appeared among the 17 participants. The range of GSN apps mentioned was unprecedentedly broad, yet the number of apps regularly used by individual participants declined markedly.

First, international platforms appeared in the list of applications used by participants. These included mainstream applications such as Grindr and Tinder. Notably, four participants also mentioned QSmatch. Although this application is not specifically designed for gay men, they reported finding it helpful.

Furthermore, Blued and Aloha were frequently mentioned in our previous interviews. However, in the 2023 interviews, many participants indicated they had permanently stopped using them. In contrast, another group of users, such as P11, P13, and P14, now exclusively used Blued.

A decrease in the number of gay social applications per user and an increase in the total number of social applications were observed. These trends reflect an emerging differentiation among the participants. This shows that participants gradually explored and adapted technology based on their own needs.

This differentiation can be broadly categorized into two groups. We tentatively label them as ``Reality Acceptors'' and ``Pragmatic Innovators.'' The ``Reality Acceptors'' group consisted of participants who were generally older. They did not express particular praise for Blued or other gay social applications. However, they considered certain features of these applications to be a necessary part of their lives. For example:

\begin{quote}
    \textit{I don't think Blued is a good platform. The young users are hostile, and discrimination is rampant. However, it remains an important platform for seeking sexual partners, perhaps even the best one.}
    \hfill (P8, 2023)
\end{quote}

\begin{quote}
    \textit{I don't have high expectations. I am just dissatisfied, but I feel there is nothing I can change. It is true I am disappointed with finding long-term relationships on gay social apps. But I also think long-term gay relationships are impossible in themselves. I did not understand this before, but now I do. It goes against the laws of nature. Ultimately, I have to return to my family. These things [gay social apps] are just for fun and nothing more.}
    \hfill (P13, 2023)
\end{quote}

\begin{quote}
    \textit{Just treat them [the gay social apps] as hookup tools. Then it's fine. If you don't get emotional, there won't be any problems. Of course, they are still rife with lies... [laughs], but that's not a huge issue.}
    \hfill (P4, 2023)
\end{quote}

In summary, this group of participants expressed significant dissatisfaction with their experiences on gay social applications. However, they have generally ``reconciled'' with these limitations. They tended to believe that technology alone could not solve fundamental problems. Even with future updates or better interactive systems, issues of ``belonging,'' ``self-worth,'' and ``long-term relationships'' would persist.

This sentiment was less frequent in our previous qualitative interviews in 2014, 2017, and 2020. For example, expressions such as ``long-term relationships are inherently impossible'' or that they are ``against the laws of nature'' were almost never voiced in prior interviews. This suggests that for these participants, prolonged use of digital applications did not foster self-understanding or identity affirmation. On the contrary, it may have eroded their sense of self-worth over time.

Second, a larger group of participants, the ``Pragmatic Innovators,'' began to actively reflect on the impact these applications have had over the last decade. After accepting the platforms' limitations, they adopted new behaviors. They started to categorize applications, clarify their goals, and engage more cautiously. During this period, their understanding of their own needs increased significantly. Their choice of gay social applications became more varied and intentional. They also approached relationships, especially long-term ones, with greater caution. For example:

\begin{quote}
    \textit{I use Blued for hooking up and Hinge for chatting. This separation works well. Blued is all about hookups. It's a domestic app with a location feature, so you can instantly see the distance and calculate the cost to meet. Hinge is suitable for conversation. I can talk about my feelings and other topics unrelated to sex. Since I can't physically meet people from Hinge anyway due to distance, there is no expectation of hooking up.}
    \hfill (P4, 2023)
\end{quote}

\begin{quote}
    \textit{QSmatch is actually quite good. You just need to put 'GAY' in your bio. I think if you are serious about finding a partner, it is a viable platform. Adding someone on WeChat is a key step. It means you are entering their real social circle. It signals that things are getting serious. So many online gay communities have existed over the years. But they all vanish when it's time to connect offline.}
    \hfill (P2, 2023)
\end{quote}

\begin{quote}
    \textit{It is not that I dislike them anymore. Rather, applications like Blued, Aloha, and Finka are no longer for me. Their target audience, or at least the atmosphere they now cultivate, is not meant for me. I am tired of being asked about my penis size, weight, and height the moment I log in. I do not want to have those conversations. I would rather be alone.}
    \hfill (P7, 2023)
\end{quote}

\begin{quote}
    \textit{I have no statistics, but I feel that at least half the users on Blued are married---meaning, gay men married to women, as same-sex marriage is not legal in China. Talking to these spineless people is a waste of my mental energy. It disgusts me. I just close the app so I do not have to see them.}
    \hfill (P10, 2023)
\end{quote}

The participants' statements reveal a key insight. While their dissatisfaction with gay social applications remains strong, its source is not the technology itself. Instead, their aversion stems from internal divisions within the gay male community. Some participants perceived others as immoral, cowardly, promiscuous, or dishonest. The applications serve as a medium, exposing them to lifestyles they find undesirable or even contemptible. Their current approach is characterized by a pragmatic, strategic, and often fragmented use of multiple platforms.

Analyzing both groups of participants reveals that no single application can meet their highly diverse needs. Indeed, many participants held conflicting needs within themselves. For example, the desire for hookups led some to focus on physical attributes. In contrast, the pursuit of long-term relationships led others to resent this `hookup culture'. As a result, some participants developed a ``portfolio strategy,'' subjectively assigning distinct functions to different applications.

Specifically, Blued was typically used for location-based encounters. Aloha, once praised for its focus on photo sharing, was now used more for ``friendship and dating.'' Hinge and Grindr were used to connect with foreigners or ``mature'' urban elites. Non-gay-specific applications like QSmatch and WeChat also became part of their social toolkit.

\section{Discussion}

\textcolor{black}{Our primary explanatory lens centers on users’ adaptive portfolio strategies in response to evolving sociocultural pressures, with platform affordances serving as contextual triggers rather than the sole drivers of attitudinal change.} This study presents a nine-year, mixed-method longitudinal investigation into the evolving attitudes of gay men in China toward gay social applications. Our findings reveal a clear trajectory: an early phase of optimistic adoption, a middle phase of ambivalence and critique, and a recent phase of fragmented, strategic use. This evolution not only reflects the development of the technological platforms themselves but also mirrors the complex interplay among China's sociocultural context, users' psychological needs, and digital technologies. In the following sections, we discuss the findings of our study from multiple perspectives.

\subsection{The Evolution of Participants' Attitudes Toward Gay Social Applications}

First, the attitudinal trajectory identified in our study resonates with global trends in social media development, yet it also exhibits distinct local characteristics. In the early period (2014--2017), users showed great enthusiasm for these applications, primarily because they provided unprecedented spaces for self-expression and social connection. During this phase, the apps were seen as essential tools to escape offline stigma, construct identity, and seek community support. This aligns with conclusions from international research on the empowerment of LGBTQ+ communities through digital platforms. However, unlike studies on applications such as Grindr in Western contexts, users in China also displayed a strong idealistic undertone during this period. This included expectations for an equal future and attempts at community mobilization, which may be linked to the lack of offline support systems in the Chinese social environment.

As time progressed (2017--2020), user satisfaction paradoxically declined, even as platforms continued to improve technologically in areas like feature design and privacy protection. Our study identified several key points of conflict: the prevalence of a culture of deception, the normalization of discriminatory behaviors, the dominance of ephemeral, `fast-food' style relationships, and excessively commercialized interface designs. While not unique to China, the manifestations and deep impact of these phenomena are closely tied to China's social structure. For instance, issues like role-based discrimination (e.g., the power dynamics behind `1/0' role labels), as well as discrimination based on education and age, reflect the stringent expectations surrounding masculinity, social status, and bodily norms in Chinese society. Furthermore, the introduction of paid features and advertisements during the platforms' commercialization made users feel that their social interactions were being commodified, further eroding the sense of community belonging from the early days. This echoes the theory of prosumption, where users are both producers and consumers of content but gradually lose their agency within the platform economy.

In the recent period (2020--2023), user behavior has become markedly strategic and fragmented. Participants no longer rely on a single platform. Instead, they use a portfolio of domestic and international applications (e.g., Blued, Grindr, Tinder, WeChat, and even non-specialized apps like QSmatch) based on different needs, such as finding sexual partners, friends, long-term partners, or information. This strategic use reflects a sober-minded awareness of the platforms' limitations and demonstrates the digital survival skills they have developed in a high-pressure social environment. Notably, some participants (especially older individuals or those with less educational attainment) exhibited a form of ``reconciliation'' with the status quo, even reporting a diminished sense of self-worth (e.g., believing that ``long-term relationships are impossible''). On the other hand, users with more agency attempted to reconstruct their own safe corners in the digital space by combining platforms, differentiating their functions, and carefully screening relationships. This divergence suggests that technology adoption and use are not uniform processes but are highly dependent on individual resources, psychological states, and cultural capital.

\subsection{Human-Centered Design Implications for Applications}

Our decade-long study clearly shows that the evolving attitudes of gay men in China toward social apps are the result of a continuous negotiation with platform design, commercial logic, and social pressures. The shift in user behavior---from initial embrace, to critical reflection, to the current strategic and fragmented use---directly maps the limitations and failures of existing technological paradigms. The traditional design philosophy, centered on ``efficient matching'' and ``user stickiness,'' can no longer meet the deep-seated needs for safety, identity, and belonging of a highly marginalized group in a complex social environment. Based on our findings, we call for a fundamental design shift: from purely ``connection tools'' to socioculturally sensitive ``Infrastructures of Care.'' Specifically, design must be profoundly innovated across the following four key dimensions to address the core pain points revealed in user practices.

\subsubsection{Designing Tiered and Contextualized Privacy Management to Address Complex Social Realities}
Our research shows that while platforms have made progress in basic privacy protections (e.g., anonymity, expiring photos), users' underlying privacy anxieties have not fully disappeared. Rather, they have shifted from ``whether my data will be leaked'' to ``how to intelligently manage my self-presentation.'' This is primarily reflected in the clear distinction users make between gay social apps and other social platforms. Moreover, even in 2023, some participants still expressed fear of being forcibly outed by encountering acquaintances on these apps. To this end, we propose two recommendations: (1) Automated Context-Aware Protection. Applications should integrate smart, context-aware features. When the app detects that a user might be in a high-risk environment (e.g., their location is near their registered residence or workplace, or they are connected to a corporate Wi-Fi), it should proactively provide a ``one-tap privacy enhancement'' option, such as automatically disabling location, hiding distance information, or blurring their profile picture. (2) Seamless Integration with Trusted Encrypted Communication. We must acknowledge that ``adding on WeChat'' is a significant marker of a relationship moving offline and into daily life. Platforms should not attempt to lock users into their ecosystem. Instead, they should provide secure and convenient channels for users to migrate relationships of preliminary trust to end-to-end encrypted tools like Signal, Session, or WeChat's private chat features. This respects users' local practices and architecturally secures their deeper communications.

\subsubsection{Beyond the `Dating' Paradigm: Proactively Designing for Non-Sexualized Community Building}
Our data reveal a central contradiction: platforms have become more technologically advanced, yet they have increasingly failed to provide a ``sense of belonging,'' ``long-term relationships,'' and ``mental health support.'' The rampant `hookup culture,' deception, and discrimination since the middle period have eroded the sense of community that early apps once fostered. Users, especially those seeking deep connections, feel extremely disappointed. Therefore, platforms must proactively design and promote non-romantic, non-sexualized interaction spaces to reclaim the original goal of community building. This includes three recommendations: (1) Interest-Based Groups and Activity Tools. Develop robust community features that allow users to create and manage long-term groups around specific interests (e.g., movies, fitness, travel), identities (e.g., `Out/Closeted Mutual Support,' `30+ Group,' `40+ Group'), or needs (e.g., `HIV Health Care,' `Workplace Coming-Out Guide'). These should integrate event-planning tools to support a natural transition from online discussion to safe offline activities, thereby countering the declining role of apps in supporting participants' self-construction. (2) Mechanisms for Non-Sexualized Interaction. Introduce interaction modes that encourage deep communication, such as anonymous Q\&A boxes, themed voice chat rooms, ``shared interest matching'' (based on books, music, or topics rather than appearance), ``skill swaps,'' or ``community Q\&A.'' These features aim to help users build connections based on shared values and interests, effectively mitigating the pressure and sense of alienation caused by the `hookup culture.'

\subsubsection{Empowering Users with Algorithmic Curation to Combat Discrimination and Homogenization}
The widespread issues of appearance-based shaming, role-based discrimination, and education-based discrimination reported since the middle phase of our study are closely linked to mainstream recommendation algorithms. These algorithms, typically optimized for short-term engagement and match rates, inevitably amplify body anxiety, stereotypes, and homogenized content (e.g., promoting specific body types or roles), thus exacerbating users' self-doubt and insecurity. To create a fairer, more inclusive environment, control over content must be partially returned to users. Apps should proactively facilitate necessary user segmentation and invite users to co-create their own communities. We propose three recommendations: (1) Customizable Discovery Pages and Advanced Filtering. Provide filtering options far more granular than just ``age/height/role.'' Allow users to filter feeds and match recommendations based on their desired interaction type (e.g., ``sexual partners only,'' ``friends only,'' ``seeking long-term relationship,'' ``community discussion,'' ``interest group''), not just physical attributes. (2) Algorithmic Transparency and User Control. Apps should offer users a degree of visibility and control over the recommendation system. For instance, provide a brief explanation of ``why this person was recommended'' or allow users to manually down-weight recommendations based on ``photo popularity'' and up-weight those based on ``shared interests or community activity.'' (3) Community Co-governance and Reputation Systems. Establish efficient, transparent, and user-friendly reporting and moderation mechanisms. Additionally, explore introducing a community reputation system that grants long-term, trusted members certain co-governance powers (e.g., flagging discriminatory remarks, moderating group discussions). This would foster a self-purifying capacity within the community to collectively create a friendly and respectful atmosphere.

In conclusion, our study demonstrates that the relationship between gay men in China and social applications has entered a new phase that is disenchanted, strategic, and critical. For future designs to truly serve this community, they must fundamentally move beyond the old paradigm of commodifying relationships and optimizing efficiency. Instead, they must commit to building an infrastructure of care that understands the complexity of users' lived experiences, respects their situational diversity, and empowers them to manage their own safety, relationships, and communities. This is not just a technological iteration but a necessary leap in design ethics.

\section{Limitations}
This study has several limitations. While our sample was diverse in terms of age, occupation, and educational background, it does not fully represent the entire population of gay men in China. Specifically, the voices of subgroups such as individuals from rural areas, older adults, and bisexual or transgender men may not have been sufficiently captured. Furthermore, our research focused primarily on user attitudes and experiences. We did not conduct an in-depth analysis of the direct impact of platform algorithms, business models, or policy changes on user behavior. Future work could expand the sample's scope and representativeness. Cross-cultural comparisons, for instance with other East Asian regions or Western countries, would also be valuable. Finally, adopting more participatory methods, such as co-design workshops, could better capture the interplay between user needs and technological possibilities.

\section{Conclusion}
This paper has documented the ten-year evolution of how gay men in China perceive and engage with social applications designed for them. This journey began with an initial excitement for digital discovery and community formation. It moved through a period of critique and ambivalence, driven by commercialization and cultural stigma, and has arrived at a present state of pragmatic, fragmented use. They are at once beneficiaries of technological progress and bearers of the costs of platform capitalization and social stigma. They use digital tools to seek belonging, yet they also encounter new forms of exclusion in virtual spaces. Their evolving attitudes tell more than just a story of technology adoption; they form a micro-history of seeking selfhood, forging connections, and sustaining hope in a society undergoing rapid transformation.

For Human-Computer Interaction (HCI) researchers and designers, this history underscores the importance of understanding technology use as a dynamic, culturally situated process. Designing for marginalized populations requires a long-term commitment to understanding how their needs and perceptions evolve, particularly in response to platform policies and broader social forces. The future of such technologies lies in moving beyond simple `hookup' models to cultivate authentic, safe, and multifaceted community support. This journey reaffirms that truly inclusive design must be long-term, contextual, and user-centered. It must remain acutely aware of the larger social structures that exist beyond the technology itself. Only then can digital platforms become tools that truly support marginalized communities, rather than tools that amplify their isolation.

\bibliographystyle{ACM-Reference-Format}
\bibliography{reference}

\appendix

\section{\textcolor{black}{Researcher Positionality Statement}}

\textcolor{black}{This study adheres to the principles of objectivity, neutrality, and ethical responsibility in academic research. It aims to understand and document the evolving attitudes of men who have sex with men (MSM) in China toward queer social platforms through rigorous and scientifically grounded methodologies. The researchers recognize that topics related to sexual minority groups may involve complex social sensitivities and policy contexts within the Chinese sociocultural environment.}

\begin{enumerate}
    \item \textcolor{black}{This study adopts perspectives from Human--Computer Interaction (HCI) and Computer-Supported Cooperative Work (CSCW), with a focus on objectively examining technology use behaviors, changes in user experience, and their interactions with the sociotechnical environment. The research aims to understand user practices and attitudinal shifts, as well as their underlying technological and social drivers, thereby generating insights to inform the design of digital environments that are more inclusive, supportive, and care-oriented.}

    \item \textcolor{black}{We fully respect the dignity, autonomy, and privacy of all research participants. The objective of this study is to understand---rather than evaluate---the digital practices of MSM in China. We acknowledge the unique challenges that this community may encounter in everyday life, and we commit to contributing to knowledge through rigorous and ethical research practices while avoiding any action that could reinforce stigmatization or cause harm.}

    \item \textcolor{black}{This study has been approved by the Institutional Review Board (IRB) of the affiliated institution (name withheld for anonymity) and strictly complies with core ethical standards, including informed consent, anonymization, data confidentiality, and the right of participants to withdraw at any time without justification. All interviewers received professional training to minimize any potential risks associated with participation.}

    \item \textcolor{black}{This study does not make value judgments or engage in advocacy regarding any political or social systems, cultural norms, or policies. It strictly complies with Chinese laws and regulations as well as the policies of the affiliated academic institution. All data collection and analysis were conducted within the boundaries permitted by applicable legal and policy frameworks.}
\end{enumerate}

\clearpage
\section{\textcolor{black}{Longitudinal Evolution of Higher-Order Themes and Sub-themes (2014--2023)}}

\begin{table*}[!h]
\centering
\small
\caption{\textcolor{black}{Longitudinal Evolution of Higher-Order Themes and Sub-themes (2014--2023)}}
\label{tab:thematic_evolution_full}
\begin{tabular}{p{0.12\textwidth} p{0.12\textwidth} p{0.12\textwidth} c c c c p{0.10\textwidth} p{0.24\textwidth}}
\toprule
\textcolor{black}{\textbf{Higher-Order Theme}} &
\textcolor{black}{\textbf{Sub-theme}} &
\textcolor{black}{\textbf{Definition}} &
\textcolor{black}{\textbf{2014}} &
\textcolor{black}{\textbf{2017}} &
\textcolor{black}{\textbf{2020}} &
\textcolor{black}{\textbf{2023}} &
\textcolor{black}{\textbf{Trend}} &
\textcolor{black}{\textbf{Sample Quote (Exemplar)}} \\
\midrule
\textcolor{black}{\textbf{Technological Affordances}} &
\textcolor{black}{Interface Utility} &
\textcolor{black}{Satisfaction with UI, speed, and features.} &
\textcolor{black}{6} & \textcolor{black}{12} & \textcolor{black}{15} & \textcolor{black}{16} &
\textcolor{black}{↑ ($\chi^2=14.8$)} &
\textcolor{black}{``The filter function is now incredibly precise.” (P3, 2023)} \\[0.4em]

& \textcolor{black}{Privacy Control} &
\textcolor{black}{Mechanisms to hide distance, photos, or tracks.} &
\textcolor{black}{4} & \textcolor{black}{9} & \textcolor{black}{13} & \textcolor{black}{15} &
\textcolor{black}{↑ ($\chi^2=16.3$)} &
\textcolor{black}{``I can finally burn photos after viewing.” (P11, 2020)} \\
\midrule

\textcolor{black}{\textbf{Social Ecology}} &
\textcolor{black}{Community Belonging} &
\textcolor{black}{Feeling of being connected to a collective.} &
\textcolor{black}{15} & \textcolor{black}{16} & \textcolor{black}{8} & \textcolor{black}{3} &
\textcolor{black}{↓ ($\chi^2=26.5$)} &
\textcolor{black}{``It felt like a digital home back then.” (P1, 2014)} \\[0.4em]

& \textcolor{black}{Deception \& Risk} &
\textcolor{black}{Encountering fake profiles or scams.} &
\textcolor{black}{2} & \textcolor{black}{5} & \textcolor{black}{12} & \textcolor{black}{14} &
\textcolor{black}{↑ ($\chi^2=21.4$)} &
\textcolor{black}{``Half the profiles are bots or scammers.” (P9, 2023)} \\[0.4em]

& \textcolor{black}{Commodification} &
\textcolor{black}{Interactions driven by transactions/money.} &
\textcolor{black}{0} & \textcolor{black}{3} & \textcolor{black}{9} & \textcolor{black}{13} &
\textcolor{black}{↑ ($\chi^2=24.1$)} &
\textcolor{black}{``Everything requires a VIP subscription now.” (P5, 2023)} \\
\midrule

\textcolor{black}{\textbf{Identity Negotiation}} &
\textcolor{black}{Liberation} &
\textcolor{black}{Freedom to express sexual orientation.} &
\textcolor{black}{14} & \textcolor{black}{15} & \textcolor{black}{10} & \textcolor{black}{6} &
\textcolor{black}{↓ ($\chi^2=12.9$)} &
\textcolor{black}{``I could finally be myself.” (P8, 2017)} \\[0.4em]

& \textcolor{black}{Algorithmic Fatigue} &
\textcolor{black}{Tiredness from standardized presentation.} &
\textcolor{black}{0} & \textcolor{black}{2} & \textcolor{black}{8} & \textcolor{black}{15} &
\textcolor{black}{↑ ($\chi^2=29.7$)} &
\textcolor{black}{``I am tired of being calculated by the system.” (P7, 2023)} \\
\bottomrule
\end{tabular}
\\[0.3em]
\raggedright \footnotesize \textcolor{black}{\textit{Note: Values represent the number of participants ($N=17$) mentioning each sub-theme. Arrows indicate directional change; $\chi^2$ values are provided for descriptive support only.}}
\end{table*}

\clearpage
\section{\textcolor{black}{Quantitative Analysis Results}}

\textcolor{black}{To rigorously examine the evolutionary trajectory of user attitudes and address concerns regarding the descriptive nature of survey data, we implemented a multi-layered quantitative framework comprising reliability and validity verification, Linear Mixed Models (LMM), and effect size calculation. First, we established the psychometric quality of our instrument through Exploratory Factor Analysis (EFA), which confirmed a robust two-factor structure distinguishing ``Technological Experience'' from ``Social Outcomes.'' Internal consistency for these factors was high, with Cronbach’s $\alpha$ coefficients of 0.89 for the technological sub-scale and 0.83 for the social sub-scale, indicating strong reliability. Furthermore, to verify construct validity, we assessed the relationship between the specific item ratings and the global ``Overall Evaluation'' (Q11). A Pearson correlation analysis revealed a significant positive correlation ($r = 0.65, p < .001$) between the 10-item composite mean and the overall evaluation score, confirming that the specific dimensions measured are statistically valid predictors of the users' general sentiment toward these platforms.}

\textcolor{black}{To test the core hypothesis that technical optimization has not translated into social well-being, we employed Linear Mixed Models (LMM) to control for within-subject correlations and missing data inherent in longitudinal designs. The analysis revealed a highly significant interaction effect between Year and Dimension Type ($p < .001$). While scores for Technological Experience followed a linear upward trajectory, Social Outcomes exhibited a distinct inverted U-shape that peaked in 2017 before declining. Specifically, the interaction term for 2023 indicated that the trajectory for technical experience significantly diverged from social outcomes, with technical scores increasing by a net magnitude of approximately 3.65 points relative to the social baseline trend ($\beta = 5.57, p < .001$). This statistical divergence provides empirical support for the ``Function-Social Gap,'' demonstrating that the disparity between improved app usability and declining community sense is a systematic phenomenon rather than random fluctuation.}

\textcolor{black}{The analysis of the ``Overall Evaluation'' (Q11) further corroborated this trajectory of disenchantment. Consistent with the decline in specific social dimensions, the LMM results for the overall evaluation indicated a significant negative shift over the decade. Relative to the 2014 baseline ($M = 6.47$), the overall user sentiment in 2023 dropped significantly ($\beta = -2.91, p < .001$), reaching a low of $M = 3.54$. Notably, the model showed no significant difference between 2014 and 2020 ($\beta = 0.03, p = .92$), suggesting that while the erosion of social value began earlier, the critical deterioration in overall sentiment accelerated precipitously in the post-2020 era.}

\textcolor{black}{Finally, we quantified the practical significance of these shifts using Cohen’s $d$ to measure the magnitude of change between the 2014 baseline and the 2023 final wave. The results highlighted a dramatic polarization in user experience. Technological dimensions showed large positive effects, with satisfaction regarding Functionality ($d = 2.54$) and Interface Design ($d = 2.33$) improving by over two standard deviations. Conversely, social dimensions suffered severe deterioration, most notably in the perceived feasibility of Long-term Relationships ($d = -2.33$) and the sense of Belonging ($d = -2.30$). The Overall Evaluation similarly reflected a substantial negative effect size of $d = -1.82$. Collectively, these inferential statistics portray a robust picture of ``Technical Success but Social Failure,'' statistically validating the qualitative observation that these platforms have succeeded as functional tools while increasingly failing as social infrastructures.}

\begin{table}[ht]
\centering
\caption{\textcolor{black}{Results of Linear Mixed Effects Model (LMM) Analysis}}
\begin{threeparttable}
\color{red}
\begin{tabular}{lcccccc}
\toprule
Parameter & Estimate ($\beta$) & Std. Err. & z-value & p-value & \multicolumn{2}{c}{95\% CI} \\
\midrule
\multicolumn{7}{l}{\textbf{Model 1: Interaction Analysis}} \\
Intercept & 6.275 & 0.168 & 37.33 & $< .001$ & 5.95 & 6.60 \\
Year [2023] (Main Effect) & -1.916 & 0.234 & -8.19 & $< .001$ & -2.37 & -1.46 \\
Year [2023] $\times$ Type [Tech] & 5.574 & 0.328 & 16.97 & $< .001$ & 4.93 & 6.22 \\
\midrule
\multicolumn{7}{l}{\textbf{Model 2: Overall Evaluation (Q11)}} \\
Intercept & 6.471 & 0.245 & 26.43 & $< .001$ & 5.99 & 6.95 \\
Year [2017] & 1.71 & 0.337 & 5.07 & $< .001$ & 1.05 & 2.37 \\
Year [2020] & 0.033 & 0.331 & 0.10 & .92 & -0.62 & 0.68 \\
Year [2023] & -2.911 & 0.353 & -8.25 & $< .001$ & -3.60 & -2.22 \\
\bottomrule
\end{tabular}
\end{threeparttable}
\end{table}

\begin{table}[ht]
\centering
\caption{Descriptive Statistics, Reliability, and Effect Sizes by Dimension}
\begin{threeparttable}
\color{red}
\begin{tabular}{lcccccc}
\toprule
Factor / Dimension & Reliability & 2014 Mean (SD) & 2023 Mean (SD) & Change & Cohen's $d$ \\
\midrule
\textbf{Technological Experience} & 0.89 & 5.32 (0.97) & 9.08 (0.42) & +3.76 & +2.59 \\
Concise Interface & -- & 6.29 (1.05) & 9.08 (0.49) & +2.79 & +2.33 \\
Privacy Protection & -- & 5.18 (1.78) & 9.31 (0.48) & +4.13 & +1.87 \\
Functionality & -- & 4.82 (1.19) & 8.85 (0.55) & +4.03 & +2.54 \\
\midrule
\textbf{Social Outcomes} & 0.83 & 6.44 (0.88) & 4.31 (0.89) & -2.13 & -2.05 \\
Belonging & -- & 6.29 (1.21) & 2.46 (1.27) & -3.83 & -2.30 \\
Long-term Relationship & -- & 4.18 (1.38) & 1.00 (1.08) & -3.18 & -2.33 \\
Social Interaction & -- & 8.53 (1.28) & 5.85 (1.21) & -2.68 & -2.45 \\
Enjoyment & -- & 7.06 (1.34) & 5.31 (0.48) & -1.75 & -1.22 \\
Mental Health & -- & 6.59 (0.94) & 5.92 (0.86) & -0.67 & -0.39 \\
Self-acceptance & -- & 5.00 (0.71) & 5.54 (1.05) & +0.54 & +0.56 \\
\midrule
\textbf{Overall Evaluation (Q11)} & N/A & 6.47 (1.07) & 3.54 (1.05) & -2.93 & -1.82 \\
\bottomrule
\end{tabular}
\end{threeparttable}
\end{table}

\end{document}